\documentclass[a4paper,11pt]{article}
\usepackage{ILD}
\usepackage{subcaption}

\usepackage{amssymb}
\usepackage{amsfonts}
\usepackage{amsmath}

\usepackage{times}
\usepackage[T1]{fontenc}

\usepackage{lineno}
\newcommand\onbehalf{\vskip 1.5in 
			 \vbox{\small
			   \hskip -0.23in The European Physical Society Conference on High Energy Physics (EPS-HEP2023)\\
 21-25 August 2023\\  Hamburg, Germany \\}}

\title{Search for new particles at the ILC}

\ildproc{phys}{2023}{015}
\date{\today}

\addauthor{M.T. N{\'u}{\~n}ez Pardo de Vera\footnote{On behalf of the IDT-WG3}}
{\institute{1}}

\addinstitute{1}{  Deutsches Elektronen-Synchrotron DESY,\\
                      Notkestr. 85, 22607 Hamburg, Germany \\
                    \vskip 0.1in
                    E-mail: maria-teresa.nunez-pardo-de-vera@desy.de}

\abstract{
  Although the LHC experiments have searched for and excluded many proposed
  new particles up to masses close to 1~\,TeV, there are many scenarios that
  are difficult to address at a hadron collider.
  This talk will review a number of these scenarios and present the expectations
  for searches at an electron-positron collider such as the International Linear
  Collider.
  The cases discussed include SUSY in strongly or moderately compressed models,
  heavy neutrinos, heavy vector bosons coupling to the s-channel in $e^+e^-$
  annihilation, and new scalars.

  \onbehalf

}

\begin{document}
\titlepage

\begin{section}{Introduction}
  The International Linear Collider, ILC, offers an excellent scenario for new particle searches.
  With respect to previous electron-positron colliders it profits from a higher luminosity and
  center-of-mass energy, polarisation of both beams, improved detector technologies and microscopic
  beam-spot. Moreover, electroweak interaction means lower background,
  allowing hermetic detectors with almost 4$\pi$ coverage and triggerless operation, and a well-known
  initial state due to the collision of point-like particles.
  Two detector concepts have been designed for the ILC, the International Large Detector, ILD, and the
  Silicon Detector, SiD. Both of them satisfy the requirements that are needed for precision Standard
  Model measurements and Beyond the Standard Model, BSM,  measurements and searches. The studies presented in
  this contribution were done using in most cases the full, Geant4 based, simulation and reconstruction 
  procedures of the ILD concept at the ILC.

\end{section}

\begin{section}{SUSY searches}
  Supersymetry is the most complete BSM theory and boilerplate for many BSM models, almost any new
  topology can be obtained in SUSY. Even if there is still no evidence of SUSY, searches for SUSY
  particles are well motivated at the ILC. In contrast to hadron colliders, the ILC offers high sensitivity
  to the electroweak SUSY sector, that is the prefered one for solving problems like naturalness,
  hierarchy, nature of dark matter or the measured magnetic moment of the muon.
  Also for scenarios with compressed spectra, pointed by many models and the global set of constraints
  from observation, the ILC is better suited than hadron colliders. 
  
  One of the most interesting channels to search for SUSY in is the direct pair-production of the
  superpartner of the $\tau$-lepton, the $\widetilde{\tau}$. The $\widetilde{\tau}$ is most likely the
  lightest of the scalar leptons, so the first one that can be found, and the
  signature of $\widetilde{\tau}$ pair production is one of the experimentally most difficult
  ones, so if the $\widetilde{\tau}$ can be seen any other scalar lepton could also be. $\widetilde{\tau}$
  searches are also motivated from the theoretical point of view, SUSY models with a light
  $\widetilde{\tau}$ could accomodate the observed relic density due to the
  $\widetilde{\tau}$-neutralino coannihilation.
  Exclusion and discovery $\widetilde{\tau}$ limits have been computed using the full simulation of the
  ILD at the ILC~\cite{stau_searches}. All SM and beam-induced backgrounds, as overlay-on-physics and only-overlay events,
  were included in the study, showing the capability to reduce them to negligible levels.
  Figure~\ref{fig:stau_higgsino}(a) shows the obtained limits and compares them with the current
  $\widetilde{\tau}$ limits coming from LEP and HL-LHC predictions, which are however highly
  model-dependent and without discovery potential for the best motivated scenarios.
  The capability of the ILC for excluding/discovering $\widetilde{\tau}$-pair production up to a
  few GeV below the kinematic limit, without model dependencies and even in the worst scenario,
  has been shown in this study.
  Higgsino searches are also well motivated since electroweak naturalness in simple SUSY models requires
  the existence of four light higgsinos, with masses around 100-300~\,GeV, in a very compressed spectrum,
  10-20~\,GeV. This scenario is again very challenge for the LHC but not for the ILC.
  Figure~\ref{fig:stau_higgsino}(b) shows the mass limits for the higgsino exclusion at the ILC with $500$\,GeV
  centre-of-mass energy. These limits were extrapolated from the LEP results taking into account only
  the polarisation of the beams and the increase in the luminosity, improvements
  due to triggerless operation at the ILC and to advance in accelerator and detector
  technologies are not considered~\cite{chargino_searches}.
  The study concludes that at the ILC either exclusion or discovery
  of Higgsinos is expected up to masses close to the kinematic limit for any mass
  difference and any mixing.
  
\begin{figure}[ttbp]
  \begin{center}
    \subcaptionbox{}{\includegraphics [width=0.49\textwidth]{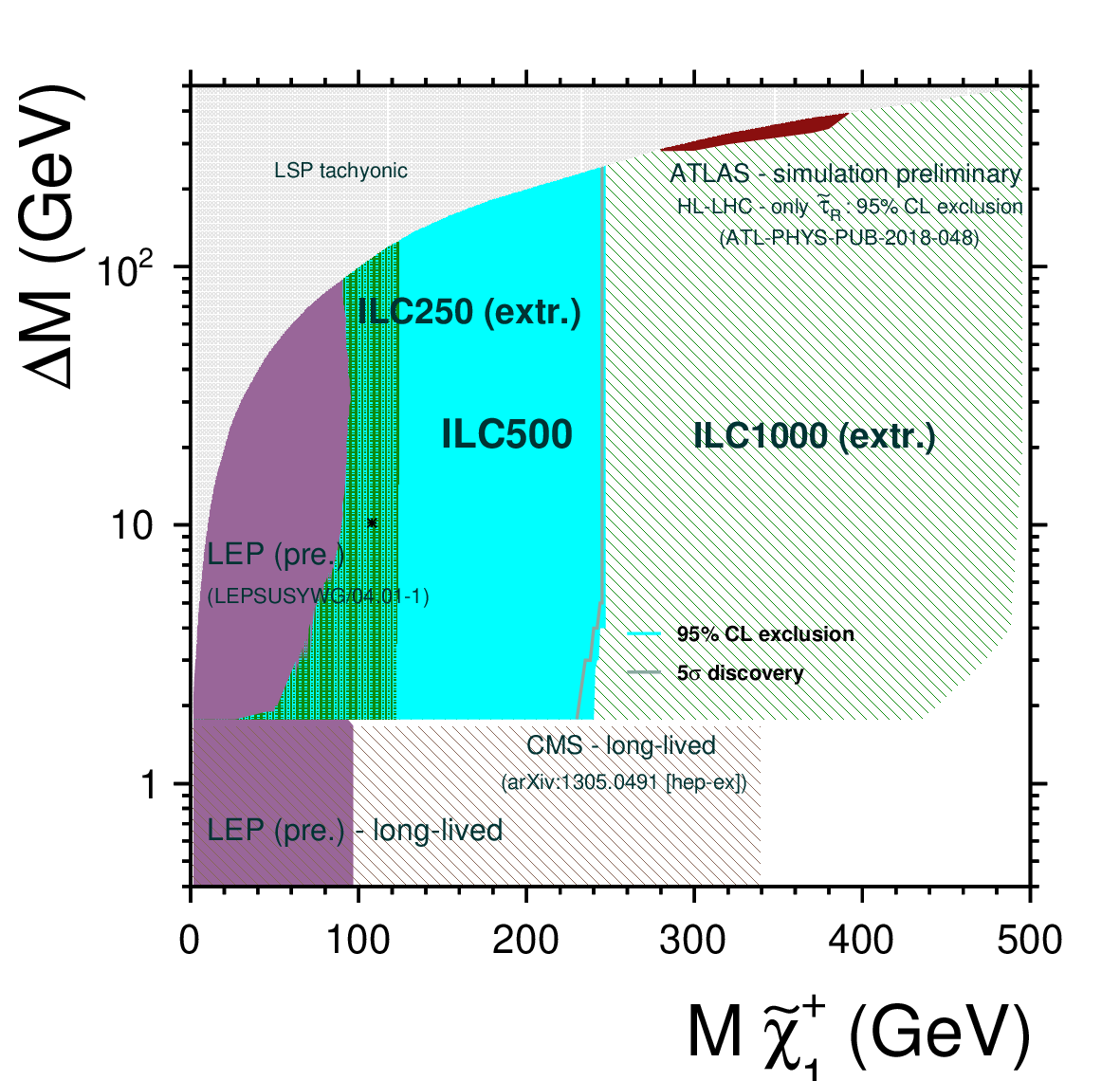}}
    \subcaptionbox{}{\includegraphics [ width=0.49\textwidth]{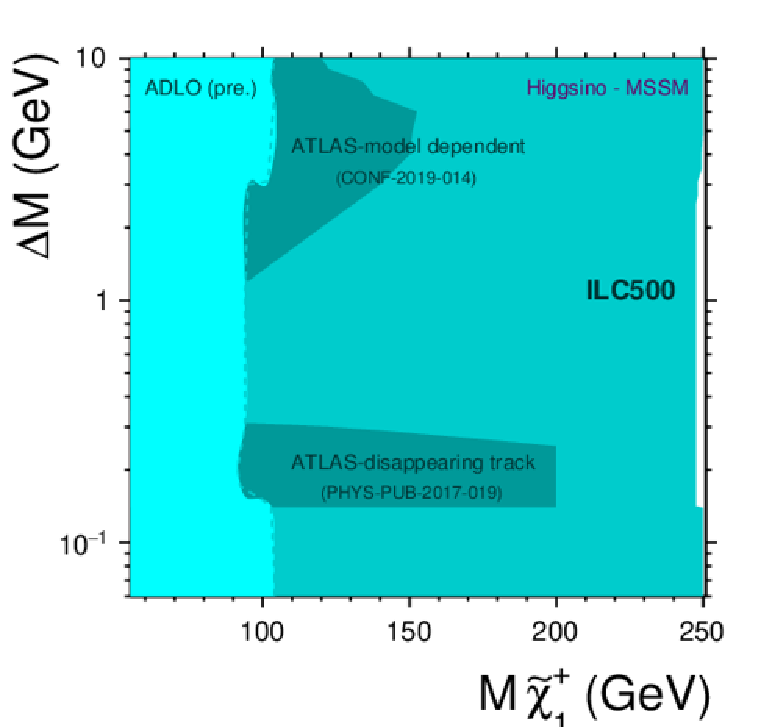}}
\end{center}
\vspace{-0.5cm}
\caption{
  (a)  $\widetilde{\tau}$ limits in the $M_{\widetilde{\tau}}$-$\Delta$M plane. ILC results are shown together with limits from LEP and projection for HL-LHC.
  The ILC region corresponds to the full H20 data-set at $E_{CMS}$ = 500 GeV, i.e. 1.6 ab$^{-1}$ at each of the beam
  polarisations $\mathcal{P}(e^{-},e^{+})$=(-80\%, +30\%) and $\mathcal{P}(e^{-},e^{+})$=(+80\%, -30\%)~\cite{stau_searches}.
  (b) ILC Higgsino mass limits extrapolated from LEP results for $E_{CMS}$ = 500 GeV~\cite{chargino_searches}.
  \label{fig:stau_higgsino}}
\end{figure}

\end{section}

\begin{section}{Searches for new Higgs-like scalars in association with a Z boson}
  Many BSM models predict the existence of Higgs-like scalars produced in association with a Z boson.
  A study searching for these new scalars at the ILC was presented in~\cite{new_scalars}. The searches were
  done for any mass of the scalar and based on the recoil of this new particle against the Z, being
  independent of its decay modes.
  The analysis used the decay of the Z boson to two muons, for that reason important detector performance
  aspects were the two muons identification and momentum reconstruction. ISR
  identification and energy reconstruction played also an important role.
  Fig.~\ref{fig:newscalars_heavyneu}(a) shows the results of this study.
  Since most of the LEP and LHC studies depend on the properties
  of the scalar, the ILC results were compared to the OPAL ones, also based on recoil against the Z.
  The results are shown for two different geometries of the ILD detector, pointing out that even if they
  have effects on the kinematical distributions used in the analysis, there is not significant difference
  in the final result. The loss of sensitivity for masses below the Z mass is due to difficulties in the ISR
  identification, being the most important limitation of the study.
  It is concluded that the ILC can exclude couplings down to a few per-cent of ones for an SM-Higgs with the
  same mass.
  The obtained limits are two orders of magnitude more sensitive than the LEP ones and cover substantial
  new phase space.

\begin{figure}[ttbp]
  \begin{center}
    \subcaptionbox{}{\includegraphics [width=0.47\textwidth]{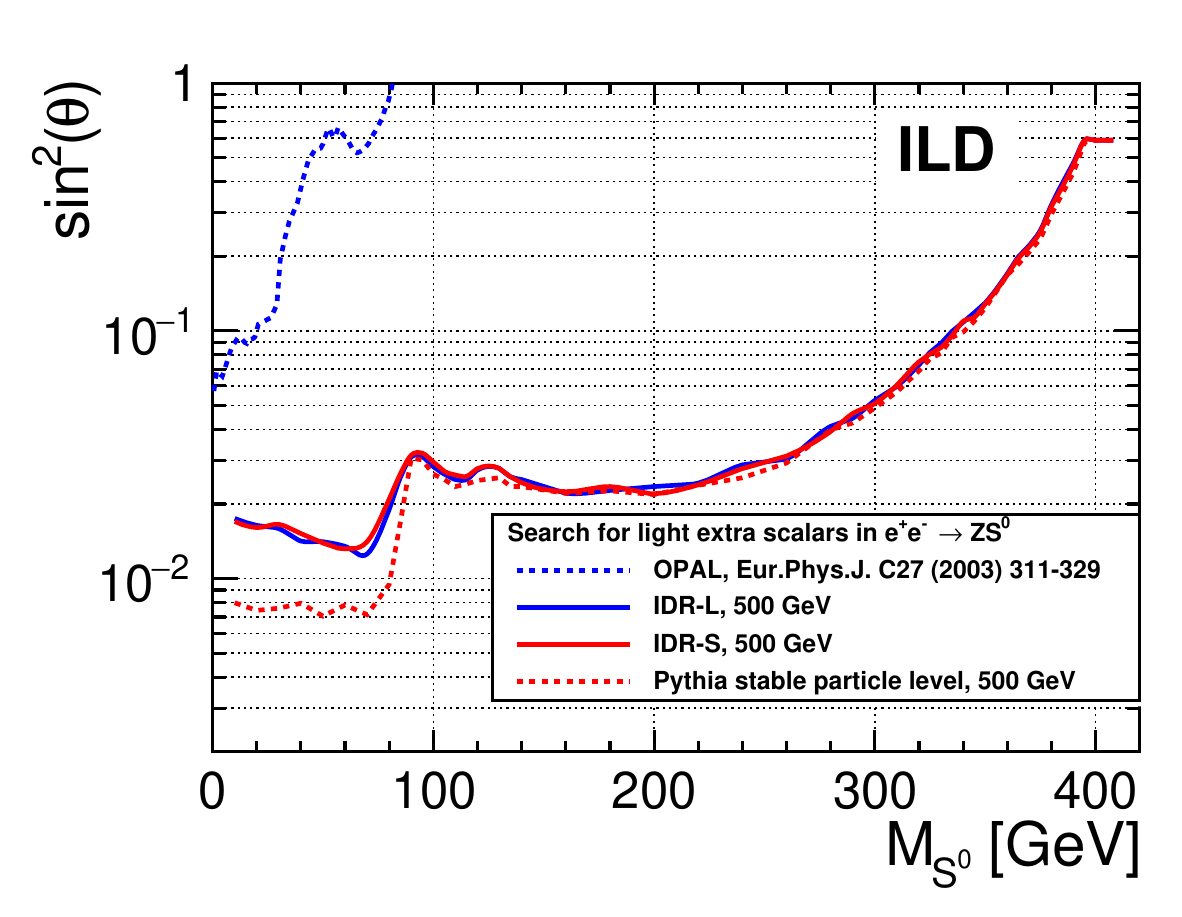}}
    \subcaptionbox{}{\includegraphics [width=0.486\textwidth]{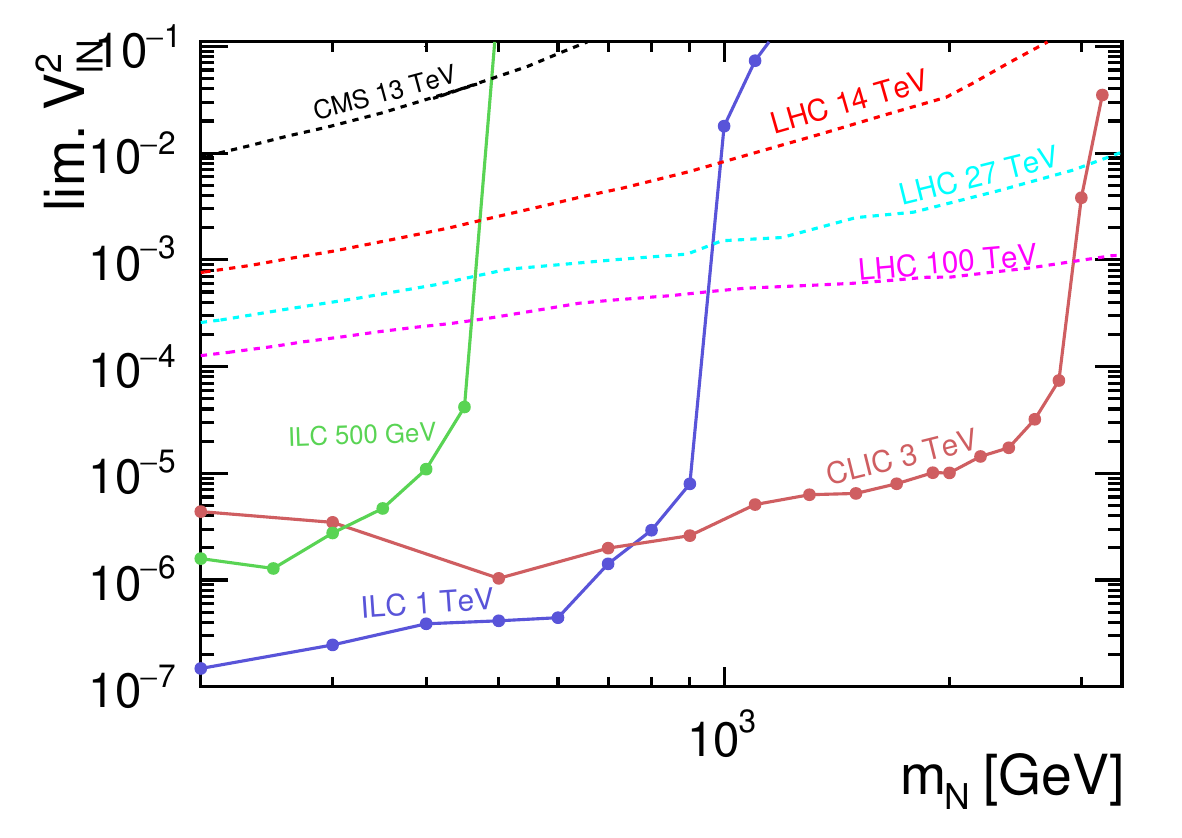}}
  \end{center}
\caption{
  (a) ILC500 projected exclusion limit for new scalars, in terms of its coupling compared to the coupling an SM Higgs at the same mass would have~\cite{new_scalars}.
  (b) Limits on the coupling $V^2_{\ell N}$ for different collider setups (solid lines: ILC500 -- green, ILC1000 -- violet, CLIC3000 -- dark red). Dashed lines indicate limits from current and future hadron colliders~\cite{heavy_neutrinos}.
  \label{fig:newscalars_heavyneu}}
\end{figure}

\end{section}

\begin{section}{Heavy neutrinos and dark neutrinos from exotic Higgs decays}

  The existence of heavy neutrinos, from either Dirac or Majorana nature, could explain many of the SM open
  problems.
  In~\cite{heavy_neutrinos}, the possibilty of the ILC for observing decays of heavy Dirac and Majorana
  neutrinos in the {\it qql} final state was studied.
  In order to make the results mostly model independent, it is assumed that only one heavy neutrino is kinematically accessible,
  allowing for flavour mixing for all three generations, and no additional gauge bosons at any
  energy scale. The study was focused on neutrino masses above the EW scale. The variables used in the analysis
  were not optimized for distinguishing between the Dirac and Majorana hypothesis, so that for on-shell production,
  the obtained expected limits were almost the same for neutrinos of both types. A further preliminary analysis
  performed by the same authors and presented in this conference uses kinematic variables that allow the
  distinction between both kind of neutrinos.
  Figure~\ref{fig:newscalars_heavyneu}(b) shows the exclusion reach as a function of the neutrino mixing parameter,
  showing that the sensitivity
  of future electron-positron colliders to the heavy-light neutrino mixing is almost independent of the heavy
  neutrino mass up to the production threshold of m$_{N}$ $\le$ $\sqrt{s}$. It is also shown that for the heavy neutrino scenarios under
  study, ILC limits are much stricter than the LHC ones and the HL-LHC prospects, being hadron colliders more sensitive to high masses and large mixings, and electron-positron colliders even better than the FCChh for much smaller mixings.

  The sensitivity of the ILC for detecting heavy dark neutrinos from exotic Higgs decays has been studied in~\cite{dark_neutrinos}.
  The process under study uses the leading Higgs production $e^+e^-\rightarrow Z^*\rightarrow ZH$, decaying the Higgs to
  a SM neutrino and a dark neutrino.
  The neutrino
  masses under study ranges between the Z and the Higgs masses. The analysis manifested the crucial role of improved jet
  clustering algorithms for future colliders. Results are shown in Fig.~\ref{fig:darkneu_indirect}(a) as a function of the
  neutrino mass and the joint branching ratio of $BR(H \rightarrow \nu N_{d}) \cdot BR(N_{d} \rightarrow lW)$, concluding
  that the exclusion limit (discovery potential) for the joint branching ratio is about 0.1$\%$ (0.3$\%$), nearly independent
  of the dark neutrino mass in the mass range under study. Interpreting these values for dark neutrino models results on
  contraints on the dark-SM mixing angles of levels down to 10$^{-4}$, a factor 10 improvement with respect to previous
  contraints.

\begin{figure}[ttbp]
  \begin{center}
    \subcaptionbox{}{\includegraphics [width=0.50\textwidth]{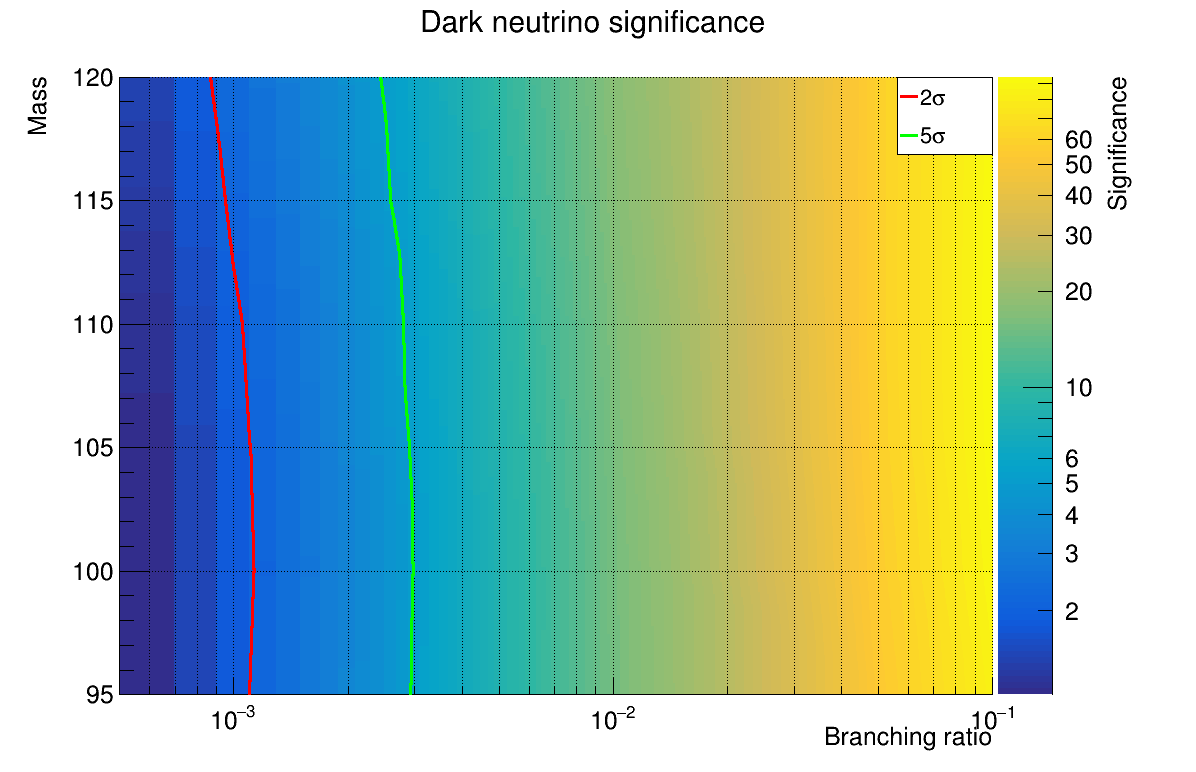}}
    \subcaptionbox{}{\includegraphics [width=0.43\textwidth]{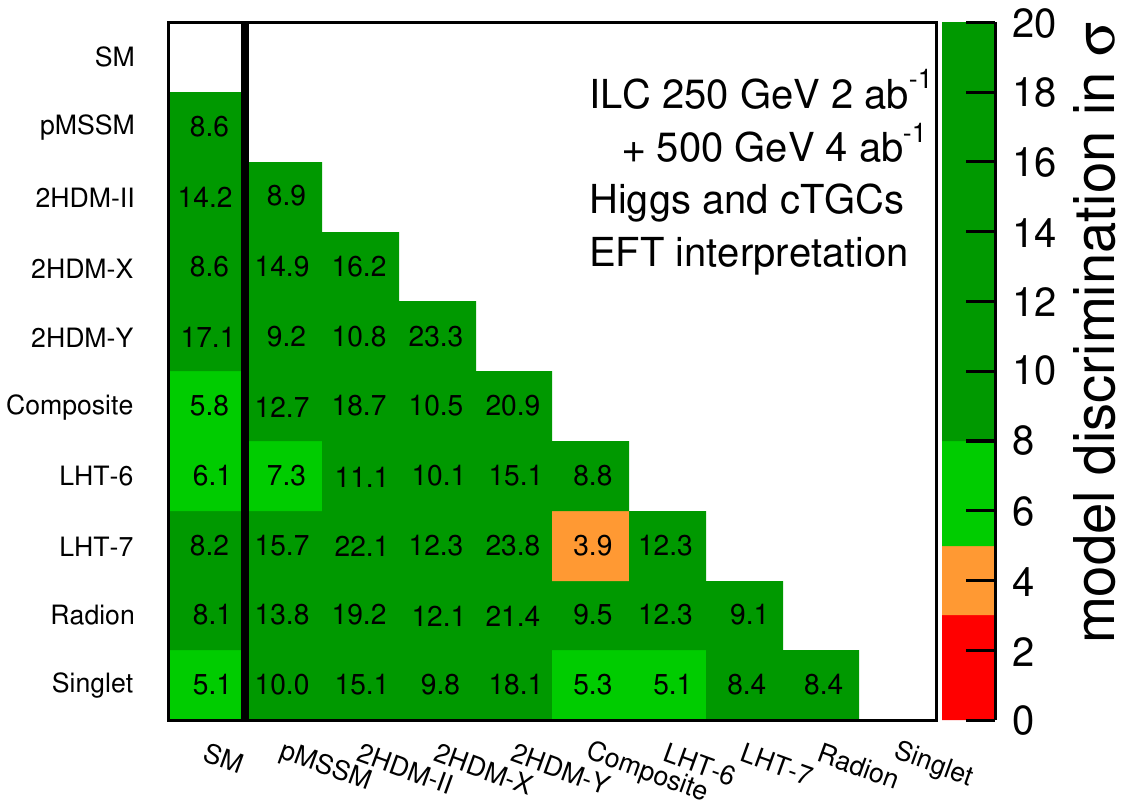}}
\end{center}
\vspace{-0.5cm}
\caption{
  (a) Dark neutrino significance as a function of branching ratio $BR(H \rightarrow \nu N_{d}) \cdot BR(N_{d} \rightarrow lW)$ and dark neutrino mass~\cite{dark_neutrinos}.
  (b) Significances of SMEFT deviations from the expectation, both for the SM expectation and the expectation of each of the various listed models~\cite{indirect_searches}.
  \label{fig:darkneu_indirect}}
\end{figure}

\end{section}

\begin{section}{WIMP Dark Matter}
  Weakly Interacting Massive Particles are among the primary candidates for Dark Matter. Searches based on simplified
  signatures, like the excess of mono-photon events, have been performed showing the sensitivity of the ILC to these kind of particles.
  The study is based on searching for an ISR photon, described within the SM and depending only indirectly on the DM
  production mechanism. Results from two simplified model aproaches~\cite{wimp_heavy, wimp_light} are shown in Fig.~\ref{fig:wimp}(a) and (b). The first one
  uses a model independent EFT approach with a heavy mediator, while the second one allows an arbitrary mediator with a sensitivity
  depending on the mediator properties. Both analyses have probed the decisive role of the ILC beam polarisations.
  The limits from these studies are stricter that the current ones also based on mono-photon signatures, which are derived from LEP
  results. The studies are complementary to those performed at the LHC.
       
\begin{figure}[ttbp]
  \begin{center}
    
    \subcaptionbox{}{\includegraphics [trim=0cm 0 0cm 0cm,width=0.49\textwidth]{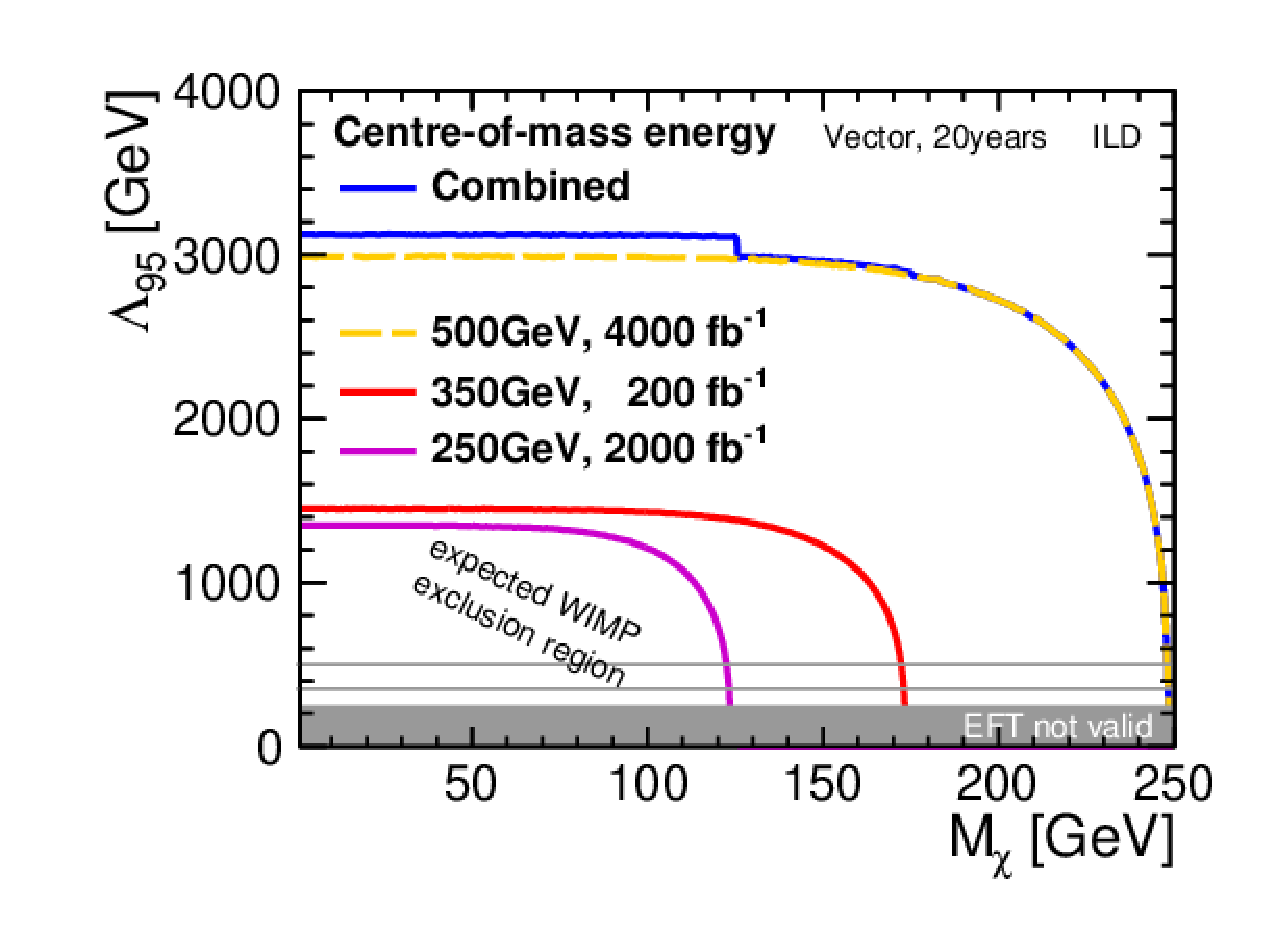}}
    \subcaptionbox{}{\includegraphics [width=0.44\textwidth]{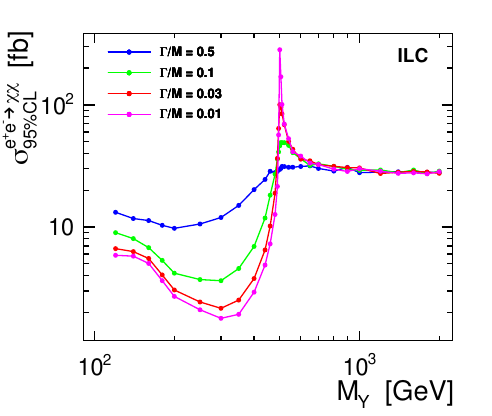}}
\end{center}
\vspace{-0.5cm}
\caption{
  (a) Dark matter searches with heavy mediators, allowing for an EFT approach~\cite{wimp_heavy}.
  (b) Dark matter searches for arbitrary mediator masses, for various assumed widths of the mediator~\cite{wimp_light}.
  \label{fig:wimp}}
\end{figure}

\end{section}

\begin{section}{Long-lived particles and indirect BSM searches}
  Long-lived particles are widely considered in many BSM scenarios and searches for new particles, like SUSY, axion-like particles, heavy neutral leptons, dark photons, exotic scalars. ILC could probe regions with small masses, couplings and mass splittings, complementary to those in which the LHC has higher sensitivity. Two challeging signatures have been studied from an experimental
  perspective, focused on searching for two tracks from a displaced vertex, with no other assumptions about the final state.
  A simple algorithm was developed offering high sensitivity in both extreme cases. Details of this study were presented in another
  contribution to this conference~\cite{long_lived}.
  
%\end{section}

%\begin{section}{Indirect BSM searches}
  Observing deviations from the behavior predicted by the SM plays an important role in not only detecting but also separating
  different BSM models.
  Figure~\ref{fig:darkneu_indirect}(b) shows results from~\cite{indirect_searches}, where SM effective field theories, SMEFT, using ILC results on
  Higgs properties and triple gauge couplings, TGCs, show significances achived to distinguish BSM model from the SM and
  between them. This study allows observing and discrimating models of new physics whose new particles are beyond the
  reach of the ILC. The authors of the study have selected models that can not be detected at the HL-LHC.

\end{section}  

\begin{section}{Conclusions}
  The potential of the ILC for direct new particle searches exceeds that of the LHC in many well-founded
  scenarios. In contrast to hadron colliders, the ILC offers clean environment without QCD backgrounds and well defined
  initial state. The ILC detectors would be more precise, hermetic and without need of trigger.
  Synenergies between ILC and LHC are expected. LHC profits from higher energy reach, while ILC is more sensitive to
  subtle signals.

\end{section}

\end{document}